\documentclass[12pt]{iopart}

\usepackage{graphicx}
\usepackage{latexsym}
\usepackage{algorithm, algorithmic}

\begin{document}         
\title{Extending the definition of modularity to directed graphs with 
overlapping communities}

\author{V. Nicosia}
\address{Scuola Superiore di Catania -- Laboratorio sui Sistemi Complessi\\
	Via S. Nullo 5/i -- Catania (ITALY)}
\ead{vincenzo.nicosia@ct.infn.it}
\author{G. Mangioni}
\address{Universit\`a di Catania -- Dipartimento di Ingegneria
  Informatica e delle Telecomunicazioni -- V.le A. Doria 6 -- Catania
  (ITALY)}
\ead{gmangioni@diit.unict.it}
\author{V. Carchiolo}
\address{Universit\`a di Catania -- Dipartimento di Ingegneria
  Informatica e delle Telecomunicazioni -- V.le A. Doria 6 -- Catania
  (ITALY)}
\ead{car@diit.unict.it}
\author{M. Malgeri}
\address{Universit\`a di Catania -- Dipartimento di Ingegneria
  Informatica e delle Telecomunicazioni -- V.le A. Doria 6 -- Catania
  (ITALY)}
\ead{mmalgeri@diit.unict.it}

\date{\today}

\begin{abstract}
  Complex networks topologies present interesting and surprising
  properties, such as community structures, which can be exploited to
  optimize communication, to find new efficient and context--aware
  routing algorithms or simply to understand the dynamics and meaning
  of relationships among nodes. Complex networks are gaining more and
  more importance as a reference model and are a powerful
  interpretation tool for many different kinds of natural, biological
  and social networks, where directed relationships and contextual
  belonging of nodes to many different communities is a matter of
  fact. This paper starts from the definition of modularity function,
  given by M. Newman to evaluate the goodness of network community
  decompositions, and extends it to the more general case of directed
  graphs with overlapping community structures. Interesting properties
  of the proposed extension are discussed, a method for finding
  overlapping communities is proposed and results of its application
  to benchmark case--studies are reported. We also propose a new
  dataset which could be used as a reference benchmark for overlapping
  community structures identification.
\end{abstract}

\maketitle

\section{Introduction}

The problem of dividing a graph into ``interesting'' subgraphs is a
classical problem in graph theory. Since graphs are used in many
application fields to represent different kinds of structures, from
relations among people to connections between computers or
interconnections among concepts, there are different reasons and
motivations for cutting graphs into smaller components. In the field
of computer networks, for example, cutting a network (graph) into
smaller components is useful to maximise network bandwidth with
respect to access to certain services, while in the case of graphs
representing communication between processes an optimal subdivision
based on minimizing the flow among group of processes is of the most
importance for optimal scheduling in multi--processors environments.

All those classical problems are often solved by different
``clustering methods'', which in turn are able to optimize a graph
structure in order to guarantee certain desired features, and a huge
amount of mathematical literature on that field has been produced.

Nevertheless, studies performed in the last few years on social and
natural networks, revealed that algorithms used for graph clustering
are neither adept at nor useful for explaining partitioning patterns observed
in those networks, such as the arising of ``communities'', ``groups''
or ``clubs''. On the other hand, those kind of structures are really
interesting, both for theoretical and for practical reasons. First,
because they naturally arise as a consequence of simple interactions
among people, and do not require complicated mechanisms to be obtained
and maintained. Second, because they have some useful properties, such
as high internal connectivity, low path length among nodes and high
robustness, which are of the most importance in real applications.

A precise definition of what a ``community'' really is does not exist
yet. One of the most widely accepted and used definitions is that
given by Newman and Girvan in~\cite{newman-2004-69}: a community
is a subgraph containing nodes which are more densely linked to each
other than to the rest of the graph or, equivalently, a graph has a
community structure if the number of links into any subgraph is higher
than the number of links between those subgraphs.

It is not so hard to accept the given definition of communities as a
reasonable one: communities in real--life are groups of strongly
connected nodes, as happens for example with people in a tennis club,
authors in a co--authorship network or colleagues working in the same
office~\cite{palla-2005-435} \cite{newman-2006-103}. It is worth
noting that usually nodes in a community know each other, and the
probability for two nodes of a community to have a neighbour in common
is higher than for other nodes in the graph
\cite{newman-2003-68}\cite{park-2003-68} \cite{brede-2005}
\cite{lusseau-2004-271} \cite{holme-2003-64}
\cite{xulvibrunet-2003-68} \cite{newman-2003}.

Properties of community structures cannot be revealed by classical
algorithms for graph clustering: those algorithms are mainly focused
on optimal subdivisions of graphs to guarantee min--flow cuts, while
finding communities requires a deeper analysis of link patterns and
relations. For this reason, a significant number of new algorithms for
community detection have been proposed in the last few years (refer to
\cite{danon-2005} \cite{fortunato-2008} for extensive overviews).

Aside from the development of algorithms for community detection, some
different metrics for community structure evaluation have been
introduced,
the most popular and widely accepted of those being the so--called
``modularity'', defined by Newman~\cite{newman-2004-69}
\cite{newman-2006-103}.  Initially defined for undirect networks, the
definition of modularity has been subsequently extended to capture
community structure in directed networks (\cite{guimera-2007-1},
\cite{arenas-2007-9}, \cite{leicht-2008}).

While an important resolution limit of the modularity measure has
been pointed out by Fortunato and Barthelemy in
\cite{fortunato-2007-104}, the modularity seems nevertheless to be a
useful measure of community structures, and many algorithms for finding
graph partitions which give optimal modularity have been proposed already 
which are able to successfully find communities in really large
complex networks\cite{PhysRevE.69.066133}\cite{clauset-2004-70}.

The only drawback of methods based on modularity optimization is that
they gives binary partitions of graphs with respect to vertices. In
other words, each vertex can be placed into just one community, and no
overlaps among communities are allowed. It is still possible to
discover sub--communities, iteratively applying those algorithms to
each of the partitions found, but discovering partially overlapped
communities is not possible at all.

On the other hand, real complex networks are never divided into sharp
sub--networks, especially those formed as a result of social
relationships and interactions: people usually belong to many
different communities, and participate to activities of a certain
number of groups at the same time.  The problem of discovering
overlapping community has been approached in the past in a very few
studies (\cite{palla-2005-435}, \cite{baumes-2005-1},
\cite{zhang-2007-1}, \cite{palla-2007-9}, \cite{lancichinetti-2008}). In this paper we
propose to extend the definition of modularity to the more general case of
directed graphs with overlapping communities.

The outline of this paper is as follows.  In section
\ref{s:newman_modularity} we give a brief description of the
modularity function as defined by Newman. In section \ref{s:overlap}
we discuss our proposal of modularity suitable for discovering
overlapped communities in directed graphs.  In section \ref{s:alggen} we present a method based on the use of a genetic algorithm to optimize modularity for overlapped communities.
In section \ref{s:results} we show results obtained maximizing the generalized
modularity of different complex networks.

\section{Newman's Modularity}\label{s:newman_modularity}

The idea behind Newman's modularity \cite{newman-2004-69} is simple: a
subgraph is a community if the number of links among nodes in the
subgraph is higher than what would be expected if links were randomly placed. This
is exactly what happens in real--world communities, where the number
and density of links among people belonging to groups (families,
clubs, user groups etc) is higher than expected in a random graph of
the same
size~\cite{newman-2003-45}\cite{newman-2003-68}\cite{park-2003-68}.

This definition of modularity implies the choice of a so--called
``null model''~\cite{newman-2004-69}, i.e. a model of graph to which
any other graph can be compared in order to assert the existence of
any degree of modularity. When testing for modularity of a complex
network, the null model used has so far been a random graph with the same
number of nodes, the same number of edges and the same degree
distribution as in the original graph, but with links among nodes
randomly placed. In such a random graph, the probability $P_{ij}$ of
having node $i$ connected to node $j$ is proportional to the degrees
(number of links) $k_i$ and $k_j$ of $i$ and $j$, respectively, and is
equal to:

\begin{equation}
P_{ij} = \frac{k_i k_j}{4m^2 }, \forall i,j
\end{equation}

where $m$ is the total number of edges in the graph. If we subdivide
an undirected graph $G(E,V)$ into a given number of subgraphs
(candidate communities), the modularity of any subgraph
$S(E',V')\subseteq G(E, V)$ with respect to the random--graph null--model 
can be computed as the sum of differences between the actual
number of links among vertices in $V'$ and the expected number of
links among those nodes in the null--model:

\begin{equation}
  Q_{S} = \sum_{i,j\in V'} \left[ \frac{A_{ij}}{2m} - P_{ij}\right]
  \label{eq:subgrmod}
\end{equation}

where $A_{ij}$ are the terms of the adjacency matrix of $G(E,V)$,
defined as:

\begin{equation}
A_{ij}= \left\{
\begin{array}{ll}
	1 & \textrm{if i and j are connected}\\
	0 & \textrm{otherwise}\\
\end{array}\right.
\end{equation}

Starting from equation \ref{eq:subgrmod}, it is possible to define the
modularity for the whole graph $G(E,V)$ as follows:

\begin{equation}
  Q = \frac{1}{2m}\sum_{i,j\in V} \left[ A_{ij} - \frac{k_i k_j}{2m
  }\right]\delta(c_i, c_j)
  \label{eq:grmod}
\end{equation}

where: 
$$
\delta(c_i, c_j) = \left\{
\begin{array}{ll}
  1 & \textrm{if i and j belong to the same community}\\
  0 & \textrm{otherwise}\\
\end{array}
\right.
$$

While this definition of modularity works only for undirected
graphs, a straightforward extension to the case of directed graphs has
been proposed in \cite{leicht-2008} \cite{arenas-2007-9}. The major
change introduced in this extended definition is that the null--model
is a directed graph as well, so that $P_{ij}$ is the probability of
having a link which starts at node $i$ and ends at node $j$, and
conversely $P_{ji}$ is the probability that a link starts at $j$ and
ends at $i$. Note that in a directed graph those two probabilities are
in general not equal, so the modularity for directed graphs is
defined as:

\begin{equation}
  Q_d = \sum_{i,j\in V} \left[ \frac{A_{ij}}{m} - P_{i,j}\right]\delta(c_i, c_j) 
  \label{eq:moddir}
\end{equation}

where

\begin{equation}
  P_{ij} =  \frac{k^{out}_i k^{in}_j}{m^2}
  \label{eq:pij_new}
\end{equation}

Note that $k^{out}_i$ is the out--degree of node $i$, i.e. the number
of links going out of $i$, while $k^{in}_j$ is the in--degree of node
$j$, i.e. the number of links coming into $j$.

This formulation of modularity, both for directed and for undirected
graphs, has been successfully used to find or confirm community
structures of a relatively large number of networks such as the
Zachary karate club network\cite{newman-2004-69}, the 
relationship network of dolphins\cite{lusseau-2004-271}, collaboration networks in
different research fields\cite{newman-2006-74}, and has proved to
catch reasonably well the structure of subgroups and communities. The
problem with this definition is that communities are sharply separated
and it does not take into account possible overlaps among communities
in the same network. 

\section{Modularity for directed graphs with overlapping communities}\label{s:overlap}

In this section we discuss our proposal to extend the 
definition of modularity to directed graphs with overlapping communities. 
We think that 
this generalization of modularity is needed in order to obtain a
metric for evaluation of real--world smooth community structures, such as
are emerging from sociological, biological and physical studies
recently made in the field of complex networks.

\subsection{Modularity for overlapping communities}
\label{subsect:mod_ov}

While overlapping among communities is an easy--to--understand
concept, since overlapping communities can be found in many real
networks, an extension of modularity to evaluate the goodness of
overlapped community decomposition is a challenging task. Looking at
how modularity was first derived by Newman in~\cite{newman-2004-69},
the first step is to choose a so--called \textit{null--model} to be
used as a reference for the definition of modularity . As reported in
section~\ref{s:newman_modularity}, Newman states that a network is
``modular'' when the actual number of connections among nodes in a
partition is higher than expected for a corresponding random graph,
where the random graph is selected as the reference
\textit{null--model}.

In the case of overlapping communities, things are a bit more
entangled, because each node can belong to many communities at the
same time, and usually it belongs to each community with a certain
strength, which is in general not equal for all communities. For this
reason, given a directed graph $G(E, V)$ and a set $C$ of overlapped
communities built from groups of nodes of $G$, an array of ``belonging
factors'' $[\alpha_{i,1}, \alpha_{i,2} \ldots \alpha_{i,|C|}]$ can be
assigned to each node $i$ in the graph, where each coefficient
$\alpha_{i,c}$ expresses how strongly node $i$ belongs to community $c$.

Without loss of generality, we can require that \

\begin{equation}
  \label{eq:alpha-limit}
  0 \le\alpha_{i,c}\le
  1 \forall i\in V, \quad \forall c \in C\
\end{equation}
and that

\begin{equation}
  \label{eq:alpha-sum}
  \sum_{c=1}^{|C|}\alpha_{i,c} = 1
\end{equation}

Note that these positions affect just the range of belonging
factors, not their meaning: a node can belong to many communities at
the same time, with different weights. The strength of each belonging
is measured as a real value in the range $[0,1]$ and the sum of all
belongings to communities is the same for all nodes in $G$.

Since each node has a belonging coefficient for each community, it is
possible to define a coefficient of belonging to each community
for edges incoming to or outgoing from a node. We can intuitively
suppose that the coefficient of belonging to community $c$ of an edge
$l = (i,j)$ which starts at node $i$ and ends at node $j$ can be
represented by a certain function of the corresponding belonging
coefficients of $i$ and $j$ to community $c$. In a formula:

\[
\beta_{l,c} = \mathcal{F}(\alpha_{i,c}, \alpha_{j,c})
\]

Definition of $\mathcal{F}(\alpha_{i,c}, \alpha_{j,c})$ is somewhat arbitrary. It
is possible, for example, to define it as the product of the belonging
coefficients of the nodes involved,  or as $\max(\alpha_{i,c},
\alpha_{j,c})$. We actually do not make any choice of a particular form
for $\mathcal{F}$.

Even without a precise idea of how a belonging coefficient
$\beta_{l,c}$ for a link $l(i,j)$ existing between $i$ and $j$ can be
derived from $\alpha_{i,c}$ and $\alpha_{j,c}$, we can still define
the null--model against which a modularity can be estimated.

Note that it is possible to rewrite equation \ref{eq:moddir} as:

\begin{equation}
Q_{d} = \frac{1}{m}\sum_{i,j\in V} \left[ A_{ij}\delta(c_i, c_j) - \frac{k_i^{out} k_j^{in}}{m
		}\delta(c_i, c_j)\right]
\end{equation}

so both the elements $A_{ij}$ of the adjacency matrix and the
probability $P_{ij}$ of having a link between $i$ and $j$ in the null
model are weighted by the belonging of $i$ and $j$ to the same
community, since $\delta(c_i, c_j)$ is equal to $1$ only when $i$ and
$j$ belong to the same community, and it is $0$ otherwise. This
formulation allows nodes to belong to only one community at a time, and
the coefficients which multiply $A_{ij}$ and $P_{ij}$ could just be
$0$ or $1$, depending on the fact the $i$ and $j$ really belong to the
same community.

Things are a bit different if we consider belonging coefficients
$\alpha_{i,c}$ as a measure of how much node $i$ belongs to community
$c$. In this case, each node can belong to many communities at the
same time, and its contribution to the modularity of a given community
should be weighted by the corresponding belonging coefficient. We can
simply reformulate modularity, where $\delta(c_i, c_j)$ is
substituted, respectively, by two different coefficients $r_{ij}$ and
$s_{ij}$, obtaining:

\begin{equation}
Q_{ov} = \frac{1}{m}\sum_{i,j\in V} \left[ r_{ij}A_{ij} - s_{ij}
	\frac{k_i^{out} k_j^{in}}{m }\right]
\end{equation}

It is also possible to put in evidence the contribution to modularity
given by each community, so that we can rewrite the modularity as:

\begin{equation}
  \label{eq:new_modul_gen}
  Q_{ov} = \frac{1}{m}\sum_{c \in C}\sum_{i,j\in V} 
  \left[ r_{ijc}A_{ij} - s_{ijc}\frac{k_i^{out} k_j^{in}}{m }\right]
\end{equation}

We can easily derive a convincing formulation for $r_{ijc}$. If
community belongings are mutually exclusive (the Newman hypothesis),
$r_{ijc} = \delta(c_{i}, c_{j}, c)$
\footnote{Note that $\delta{c_i, c_j, c}$ is equivalent to
  $\delta{c_i, c_j}$, where we include the community index
  $c$ only to make it consistent with the formulation of modularity
  which takes into account contributions given by each community.}  is
the portion of the contribute to modularity given by community $c$ due
to link $l(i,j)$ and this portion is equal to $1$ if and only if $i$
and $j$ are both into the same community and this community is exactly
$c$, i.e. if and only if $c_i = c_j = c$, otherwise it is equal to
$0$. If we think of $r_{ijc}$ as the weight of the contribution of
$l(i,j)$ to modularity of community $c$, we can define it, in the case
of overlapping communities, as the belonging coefficient of $l(i,j)$
for community $c$:

\begin{equation} 
  r_{ijc} = \beta_{l(i,j),c}  = \beta_{l,c} = \mathcal{F}(\alpha_{i,c}, \alpha_{j,c})
\end{equation}

A neat definition of $s_{ijc}$ is a bit more complicated, and requires
a clear definition of the \textit{null--model} to be used as
reference.  We observed that in graphs which have a significant
modularity, the modularity of a partition is measured as the
difference between the number of links which are ``internal'' to each
community and the number of total links originated by nodes in the
partition. This means that for modular graphs the probability that two
nodes belong to the same community is higher if those nodes are
neighbours. For this reason, a suitable null--model could be a random
graph without a community structure, where the probability for a node
to belong to any partition is not related to the fact that any another
node belongs to the same partition.

Putting it in a clear way, given a graph $G(E,V)$ we choose as
null--model a random graph corresponding to $G(E,V)$ where each node
has an out--degree and in--degree as in the original graph, and where no
particular community partition can be derived by structural properties
of the graph, i.e. where the probability that a node $i$ belongs to a
given community $c$ with a belonging factor $\alpha_{i,c}$ does not
depend upon the probability that any other node $j$ in the network
does belong to the same community with $\alpha_{j,c}$.  The latter
condition is equivalent to saying that the expected belonging
coefficient of any possible link $l(i,j)$ starting from a node into
community $c$ is simply the average of all possible 
coefficients of belonging to $c$ of $l$, so:

\begin{equation}
  \beta_{l(i,j),c}^{out} = \frac{\sum_{j\in V} \mathcal{F}(\alpha_{i,c},
    \alpha_{j,c})}{|V|}
\end{equation}

Accordingly, the expected belonging coefficient of any link $l(i,j)$
pointing to a node going into community $c$ is:

\begin{equation}
  \beta_{l(i,j),c}^{in} = \frac{\sum_{i\in V} \mathcal{F}(\alpha_{i,c},
    \alpha_{j,c})}{|V|}
\end{equation}

Those belonging coefficients are used to weight the probability of
having, respectively, a link starting at node $i$ and a link pointing
to node $j$.  Modularity in the case of overlapped communities can be
accordingly formulated as:

\begin{equation}
  Q_{ov} = \frac{1}{m}\sum_{c \in C}\sum_{i,j\in V} \left[
    \beta_{l(i,j),c}A_{ij} -
    \frac{\beta_{l(i,j),c}^{out}k_i^{out}\beta_{l(i,j),c}^{in}k_j^{in}}{m}\right]
\end{equation}

Note that the extension of the definition of modularity to the case of
overlapping communities still depends on the choice of
$\mathcal{F}(\alpha_{i,c}, \alpha_{j,c})$, i.e. on the way we choose
to weight the contribution of each edge to the modularity calculated
for community $c$.  

The most important properties required for modularity in the
formulation given by Newman were that:

\begin{enumerate}
\item
  \label{cond:1}
  $Q = 0$ when all nodes belong to the same community, i.e. when no
  community structure can be inferred from topological considerations
\item
  \label{cond:2}
  Higher values of $Q$ indicate stronger community structure
\end{enumerate}

Both of these conditions are satisfied by $Q_{ov}$ as well: if all nodes
belong to the same community (condition \ref{cond:1}), $|C| = 1$ and 

\[
\forall i \in V \alpha_{i,1} = 1
\]

as required in order to satisfy equation \ref{eq:alpha-sum}. At the
same time $\mathcal{F}(\alpha_{i,c}, \alpha_{j,c})$, which expresses a
coefficient of belonging to the unique community $c$ for the edge that
links $i$ and $j$, should give a value of $1$ for $l(i,j) \forall i,j
\in V$, and this condition is fully satisfied, for instance, when
$\mathcal{F}(\alpha_{i,c}, \alpha_{j,c})$ is a simple average among
$\alpha_{i,c}$ and $\alpha_{j,c}$, or if it is the product of the two
values or the max between the two. This implies that

\begin{equation}
  \beta_{l,c}^{out} = \frac{\sum_{j\in V} \mathcal{F}(\alpha_{i,c},
    \alpha_{j,c})}{|V|} = \frac{|V|}{|V|} = 1
\end{equation}

and, similarly: 

\begin{equation}
  \beta_{l,c}^{in} = \frac{\sum_{i\in V} \mathcal{F}(\alpha_{i,c},
    \alpha_{j,c})}{|V|} = \frac{|V|}{|V|} = 1
\end{equation}

so $s_{i,j} = 1$ and the modularity simply reduces to:
\begin{equation} 
  Q_{ov} = \frac{1}{m}\sum_{i,j\in V}\left[A_{i,j} - \frac{k_i^{out}
      k_j^{in}}{m}\right] = 0
\end{equation}

because
\[
\sum_{i,j\in V} A_{i,j} = m
\]
and
\[
\sum_{i,j\in V} \frac{k_i^{out}
  k_j^{in}}{m} = m
\]

So condition (\ref{cond:1}) is satisfied. At the same time, condition
\ref{cond:2}) is satisfied in a network with a modular structure,
since the second member of $Q_{ov}$ represents the case of a
completely unstructured network, where even the coefficient
of belonging to a given community $c$ of a node $i$ is absolutely unrelated with
the coefficients of belonging to the same community of nodes in the
neighbourhood of $i$. This is clearly not the case in real complex
network: it is much more probable that a node's coefficient of belonging to
a given community is similar to the coefficients of belonging to the same
community of its neighbours.  So a strong (overlapping) community
structure does imply higher values of $Q_{ov}$, thus satisfying
condition (\ref{cond:2}).

\section{Modularity optimization as a genetic problem}\label{s:alggen}

In the last three decades genetic algorithms (GAs) have played an
important role in the field of optimization techniques. Such
algorithms are especially suitable when the solution space of a given
problem is very large, and an exhaustive search by brute--force for
the optimal solution is practically unfeasible.

GAs are mainly based on simulating the
creation and evolution of a population of individuals, characterised
by a {\em chromosome}, where each individual represents a possible
solution of the optimization problem. At each simulation step, better
individuals are selected, included in a new generation and used to
create new elements which replace the worst individuals of the
previous generation. The convergence toward a satisfactory solution is
obtained applying a set of proper (problem-dependent) operations to
each individual.  At each iteration, the fitness of all population
members is evaluated, and individuals are ordered on the basis of
their {\em fitness} level. The individuals having the highest fitness
value are replicated in the next generation, while new individuals are
created, combining the best members of the previous generation using a
{\em crossover} operation, as well as by performing random {\em
  mutations}.

The most interesting characteristic of GAs is the low computational
complexity, while the main drawback is the risk of finding non-optimal
solutions, since the optimisation procedure can be stalled by local
maxima.  

A detailed description of GAs is beyond the scope of this paper. On
the other hand we are interested in using GAs for modularity
optimization: since $Q_{OV}$ can be considered a fitness function
(better decompositions of the graph correspond to higher values of
$Q_{ov}$), it is necessary to map the modularity optimization problem
into a genetic problem, where the fitness function is $Q_{ov}$ itself
and solutions are represented by possible partitions of a graph into
overlapping communities. A similar approach for the simple case of
maximizing classical modularity has been proposed by Bingol et al. in
\cite{gunes-2006}.

\subsection{Chromosome representation}
\begin{figure}[!htb]
	\centering
	\includegraphics[scale=0.6, keepaspectratio]{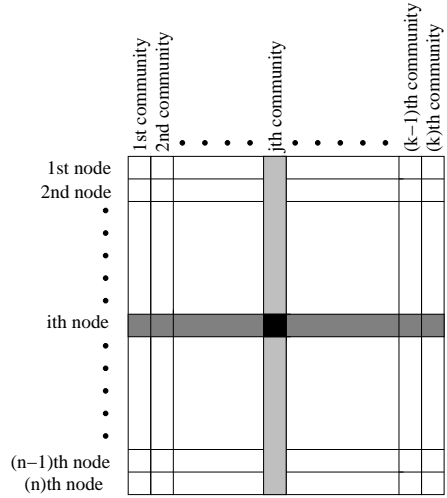}
	\caption{Chromosome representation}
	\label{fig:chromo}
\end{figure}

First of all we need to find a chromosome representation suitable for
the given problem.  In our implementation the chromosome is
represented by a matrix $M=(\alpha_{i,c}), \;
where\;i=1,...,|V|\;and\;c=1,...,|C|$. Each element $\alpha_{i,c}$ is
the strength with which a graph node $i$ belongs to a community $c$. Note that
$\alpha_{i,c}$ ranges in the interval [0.0,
  1.0]. Figure~\ref{fig:chromo} shows a graphic representation of the
chromosome.

As discussed above, each node of the graph is subject to the following constraint:   
\begin{equation}
	\sum_{c=1}^{|C|} \alpha_{ic} = 1.0
	\label{eq:norm}
\end{equation}

Equation (\ref{eq:norm}) represents a normalisation to 1.0 of node 
factors of belonging to the communities. Such a constraint is used to avoid the
GA converging toward solutions in which nodes belong to several communities at the same
time with a factor 1.0. In other words, this means that we are interested in 
finding solutions for which a node having a unity belonging factor can only
belong to one community.

\subsection{The Algorithm}\label{ss:alg}

\begin{algorithm}[hbt]
\footnotesize
\caption{Genetic algorithm main function}
\label{alg:main}
\begin{algorithmic}[1]
	\REQUIRE  $n\_epochs$, $n\_individuals$, $n_{comm}$
	\STATE $population = CreatePopulation(n\_individuals, n_{comm})$
	\FOR{($i=0$ to $n\_epochs$)}
	\STATE $population.Fitness ()$
	\STATE $population.SortingAndSelection ()$
	\STATE $population.CrossOver()$
	\STATE $population.Mutation ()$
	\STATE $population.CleanUp ()$ 
	\ENDFOR
\end{algorithmic}
\end{algorithm}
\label{fig:algo}

As shown in algorithm \ref{alg:main}, the algorithm implemented to
discover communities takes as inputs three parameters: $n\_epochs$,
i.e the total number of generations (i.e. the number of simulation
cycles); $n\_individuals$, which is the number of individuals making
up the population; $n_{comm}$, the maximum number of overlapped
communities that the algorithm will try to find. The last mentioned
parameter is not mandatory, but it is useful for reducing the
size of the solutions space and to avoid large chromosomes.

The first step of the GA is the
creation of the initial population of $n\_individuals$ elements.  At
this step, the chromosome of each individual is initialized with
belonging factors chosen at random in the range $[0.0, 1.0]$ and
normalized in order to respect the constraint expressed by equation
(\ref{eq:norm}). 
After initialization, the GA runs for $n\_epochs$ steps, applying at
each iteration a set of genetic operators on the population
individuals.  More specifically, fitness evaluation, selection,
crossover and mutations operations are performed on the individuals,
as explained in detail in the following.

\subsubsection*{Fitness evaluation, sorting and selection.}
Fitness evaluation consists in calculating a modularity function for
each individuals chromosome. As stated previously, a given number of
individuals with higher fitness are included in the next
generation. In order to preserve the number of individuals, we need to
add new members. They are created as follows: a given number of them
is obtained from crossover among better individuals of the previous
generation, while the remaining members are created from scratch. In
summary, the new generation contains both the best and a combination
of better elements of the previous generation.

\subsubsection*{Crossover and Mutation.}
Crossover is a genetic operation which consists on exchanging a
portion of the chromosomes of two different individuals. The aim of
this operation is creating a new individual which inherits its genetic
structure (i.e. its chromosome) from the ones of two other population
members, hoping that the new individual is better than those it derives
from.

\begin{figure}[!htb]
	\centering
	\includegraphics[scale=0.48, keepaspectratio]{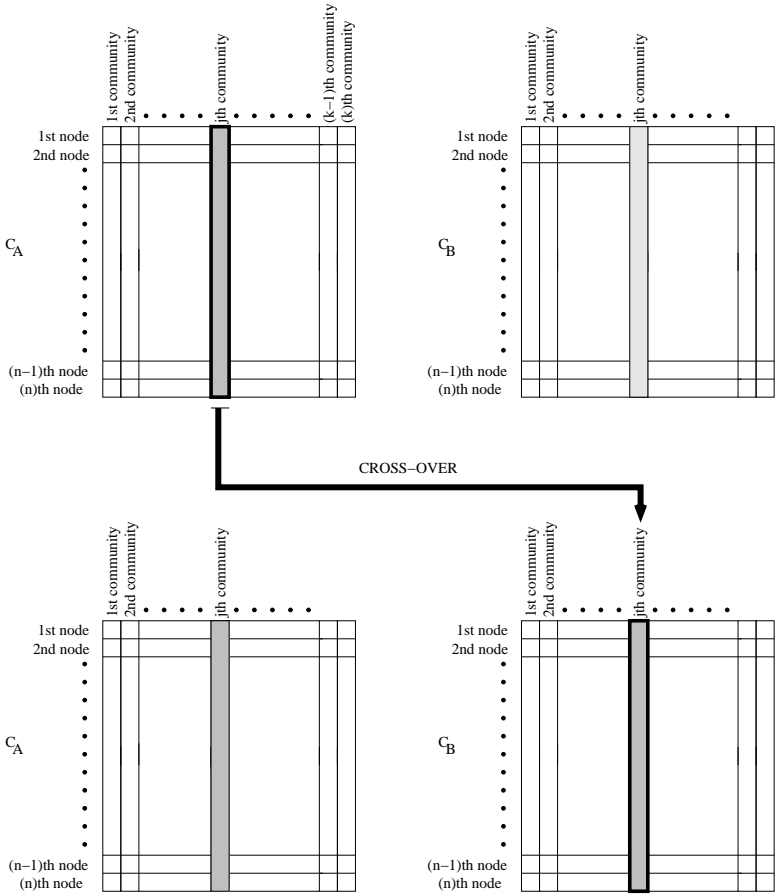}
	\caption{Cross--over operation}
	\label{fig:crossover}
\end{figure}

In the past, several different crossover techniques have been proposed. In
figure~\ref{fig:crossover} the crossover operation
that we have used in our GAs is graphically represented.
This crossover operation takes as inputs chromosomes $C_A$ and
$C_B$, respectively belonging to individuals $A$ and $B$; it works as
follows: 1) a random number $j$ in range $[1, K]$ is chosen (i.e. a
community is randomly selected); 2) the values in column $j$ of $C_A$
are copied into column $j$ of $C_B$. This crossover defines a one--way
operation, since there are values exchanged in one direction only, from
$C_A$ to $C_B$. After a crossover, $C_B$ is modified while $C_A$
structure remains unchanged.

Moreover, in order to improve the chances of finding the optimal
solution, at each iteration mutations are applied on a set of
individuals. We have defined a mutation operation that randomly
selects a given number of individuals changing in their chromosomes
the factor of belonging to a community for a node, where both node and
community are chosen at random.   
 
\subsubsection*{Clean--up.}
Besides the standard genetic operations, in the proposed GA we have
also implemented a $CleanUp$ function with the aim of improving the
quality of the graph partition into overlapping communities. Such a
function is applied to the chromosome of each individual and it is
strictly related to the optimization problem that we are dealing with.  In
particular, for a given chromosome, a node $i$ and a community $c$ are
randomly selected and the average belonging factor $avgNeigh(i,c)$ for 
community $c$ for neighbours of $i$ is computed (see
equation(~\ref{equ:avgNeigh})).  Moreover, the average belonging factor
$avgNotNeigh(i,c)$ for community $c$ of the nodes that are not
neighbours of $i$ is evaluated (see equation(~\ref{equ:avgNotNeigh})).

\begin{equation}
	avgNeigh(i,c) = \frac {\displaystyle \sum_{v \in Neighborhood(i)}
	  \alpha_{v, c}}{|Neighborhood(i)|}
	\label{equ:avgNeigh}
\end{equation}

\begin{equation}
	avgNotNeigh(i,c) = \frac {\displaystyle \sum_{v \notin Neighborhood(i)}
	  \alpha_{v, c}}{n - |Neighborhood(i)|}
	\label{equ:avgNotNeigh}
\end{equation}

The values of $avgNeigh(i,c)$ and $avgNotNeigh(i,c)$ are then
compared: if the former is greater than the latter, the belonging
factor of node $i$ for community $c$ is increased by a given small
amount; for the contrary, it is decreased.  The reason why $cleanUp$ is
effective can be explained noting that the neighbour of node $i$ has,
on average, a higher probability of belonging to community $c$ than the
nodes that are not in neighbourhood of $i$. We make use of this
information to drive the algorithm towards a meaningful solution
(i.e. towards the optimum) by further increasing the belonging factor
of node $i$ for community $k$ to speed up convergence. On the other
hand, (i.e. $avgNeigh(i,c) < avgNotNeigh(i,c)$), decreasing the
belonging factor of node $i$ for community $c$ is a way to take into
account that node $i$ probably should belong to $c$ less strongly.

In our experiments, we found that the clean--up function significantly
improves both the quality of community partitions and the speed of the
GA convergence.

After crossover, mutation and clean--up operations, each chromosome could not 
respect normalization constraint expressed by equation (\ref{eq:norm}).
For this reason, a normalization operation is performed on each chromosome
before starting with a new iteration.

\subsubsection*{Computational complexity of the algorithm.}
\label{sect:GA:computational}
In order to evaluate the computational complexity of the proposed
algorithm, we will only take into account operations depending on the
size of the input data, i.e. the number $n$ of network nodes and the
number $|C|$ of communities.  Analysing the algorithm reported in
subsection \ref{ss:alg}, it is possible to note that the most critical
operation from the computational complexity point of view is the
fitness evaluation. Studying this function, we can conclude that the
proposed algorithm has a complexity of $O(|C|*n^2)$ in the worst case.

\section{Results}\label{s:results}

In this section we show results obtained maximizing the generalized
modularity of different complex networks, and precisely:

\begin{itemize}
\item 
  The Zachary Karate Club
\item
  Social relationships of dolphins
\item
  The network of reviews of political books
\item
  The network of students at engineering faculty
\end{itemize}

The first two networks are well--known to researchers interested in
finding community structures, because they have been widely used as
benchmarks for testing the quality of new algorithms for community
detection. The third network is obtained by connecting together about
one hundred political books according to other books that they have been
sold together with. Each book has attached a label, corresponding to the
political party the author belongs to, according to reviews given by
readers. The fourth network is a new dataset built as a complement to
this work, obtained by asking students in a computer engineering class 
at the University of Catania which colleagues they know the
most. Following sections describe those networks in depth and report
results of $Q_{ov}$ optimization.

As stated in section \ref{subsect:mod_ov}, it is necessary to choose
$\mathcal{F}(\alpha_{i,c}, \alpha_{j,c})$, which expresses the
belonging to community $c$ of a link connecting node $i$ to node $j$
as a function of the coefficients of belonging  to
community $c$ itself of $i$ and $j$.

In our experiments we tried several different bidimensional functions,
but the best results so far were obtained so far when $\mathcal{F}$ is a
two--dimensional logistic function:

\begin{equation}
  \mathcal{F}(\alpha_{i,c}, \alpha_{j,c}) = 
  \frac{1}{(1 + e^{-f(\alpha_{i,c})})(1 + e^{-f(\alpha_{j,c})})}
\end{equation}

where $f(\alpha_{i,c})$ is a simple linear scaling function:

\begin{equation}
  f(x) = 2px - p, p \in \mathcal{R} 
\end{equation}

Note that $\mathcal{F}(\alpha_{i,c}, \alpha_{j,c})$ is
\textit{practically} zero when both $\alpha_{i,c}$ and $\alpha_{j,c}$
are equal to zero, and it is \textit{practically} 1 when both
$\alpha_{i,c}$ and $\alpha_{j,c}$ are equal to 1, so that condition
(\ref{cond:1}) is respected. The choice of a two--dimensional logistic
function for $\mathcal{F}$ is due to the fact that it is a fairly
smooth non--linear version of the $\delta$ distribution used in the
traditional modularity definition, and gives reasonable values for
different compositions of $\alpha_{i,c}$ and $\alpha_{j,c}$.

\subsection{The Zachary Karate Club}

\begin{figure}[!htbp]
  \begin{center}
    \includegraphics[scale=0.5]{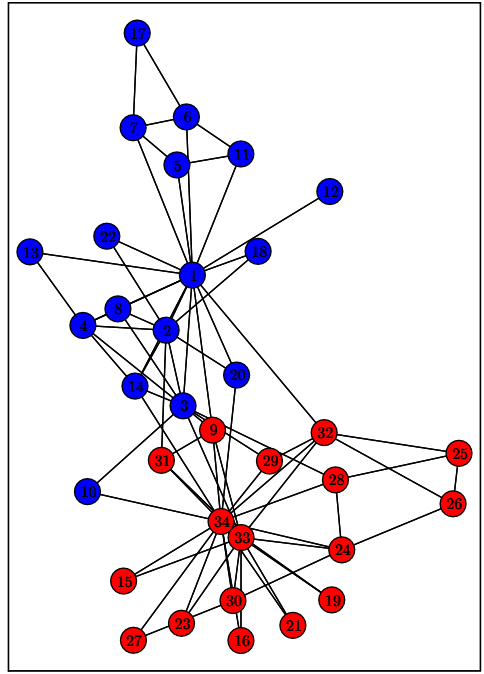}
  \end{center}
  \caption{The two main communities of the Zachary Karate Club
    Network}
  \label{fig:zachary2}
\end{figure}

The Zachary Karate Club is a social network formed by relationships
among people in the same karate club which has been extensively used
as a benchmark for all algrithms which aim to discover communities in
complex networks. Many algorithms for community detection found the
correct partition of this network into the two main communities, as
showed in figure \ref{fig:zachary2}.

\begin{figure}[!htbp]
  \begin{center}
    \includegraphics[scale=0.5]{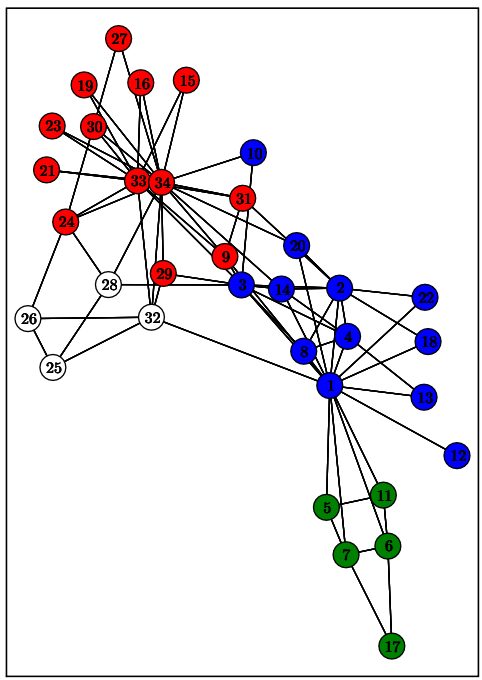}
  \end{center}
  \caption{The four communities found maximising the modularity of the
    Zachary Karate Club Network}
  \label{fig:zachary4}
\end{figure}

Nevertheless, further studies on this network showed
\cite{duch-2005-72} that if the constraint of having just two
communities is relaxed and we search for partitions with more
communities, then there is a better decomposition which has a higher
modularity value and finds four communities. In particular, the two
main communities are further divided into two sub--communities; the
real decomposition of the Zachary Karate Club is depicted in figure
\ref{fig:zachary4}.

Using GAs for optimization of $Q_{ov}$ for the Zachary network, and
fixing the number of communities to be found (as
mentioned in section \ref{sect:GA:computational}) at 2, then the best
decomposition obtained by the GA finds the maximum modularity when
$\alpha_{3,0}=0.81$ and $\alpha_{10,0}=0.63$, while $\alpha_{i,0}
=1.0$ or $\alpha_{i,0} = 0.0 \forall i \notin [3,10]$, according to
the correct placement of nodes into the community in which they belong.

\begin{figure}[!htbp]
  
  \begin{center}
    \includegraphics[scale=0.3]{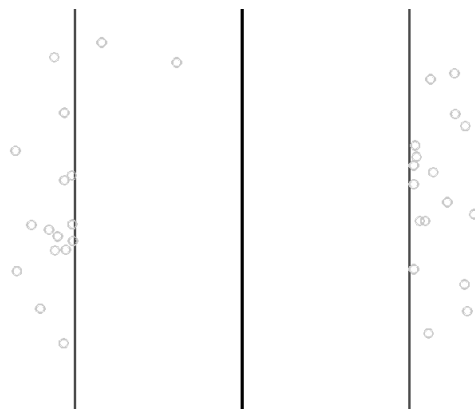}
  \end{center}
  \caption{Overlap among the two communities in the Zachary Karate
    Club -- optimal solution}
  \label{fig:zachary2_ov1}
\end{figure}

The overlap among the two communities is shown in figure
\ref{fig:zachary2_ov1}. The two lateral lines indicate community
boundaries: all the nodes which are to the left of the leftmost
vertical line belong only to the first community, and all nodes to the
right of the rightmost vertical line belong only to the second
community. The vertical line in the middle of the graph corresponds to
a perfect overlap, i.e. if a node stands on that line than it belongs
50\% to the first community and 50\% to the second. 

While the solution reported in figure \ref{fig:zachary2_ov1} is
found by the GA in 98\% of the runs, a couple of other solutions which
have almost the same modularity value sometimes come out as local
maxima. These solutions discover a couple more nodes as overlapped
among communities, even if the belonging coefficients for their original
community are really close to 1. These solutions are interesting in 
pointing out that it is important to take into account not only the
partition which has ``the best'' modularity, but also other partitions
which are sub--optimal but still meaningful.

\begin{figure}[!htbp]
  
  \begin{center}
    \includegraphics[scale=0.3]{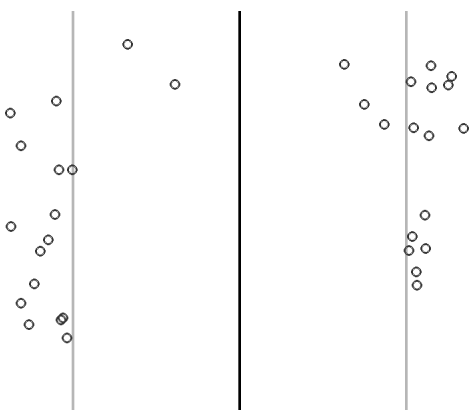}
  \end{center}
  \caption{Overlap among the two communities in the Zachary Karate
    Club -- sub--optimal solution}
  \label{fig:zachary2_ov2}
\end{figure}

In figure \ref{fig:zachary2_ov2} one of these sub--optimal solution is
reported. See how the best solution is somehow included in this one,
since node 3 and 10 are overlapped with almost the same belonging
coefficients as in the best case, while other overlaps appear in the
other communities: those nodes are nodes 9, 31 and 34, and all of them
are in the border of their community, and have links to nodes in the
first community as well.

After having constrained the GA to explore the space of solutions 
in the case of a partition with only two communities, 
we tried to maximise modularity raising the
upper bound of communities to be found by the GA to ten. The
surprising result was that six of the ten available communities were
left completely empty by the algorithm, and just four of them contain
nodes: two of them correspond to the two original communities found by
all algorithms, with the overlaps of node 3 and 10, while the other
two communities are the two sub--communities found in
\cite{duch-2005-72} massively overlapped with the main
communities. This result clearly shows that the generalised
formulation of modularity is not only able to capture overlaps but, to
some extent, also to take into account hierarchical organisation of
communities.

\subsection{The social network of dolphins}

Nodes in a social networks do not necessarily have to be humans, and
links among nodes in such networks are not limited to friendship. An
extension of the concept of social network is possible also for groups
of animals, where links usually express some kind of relationships
among them, e.g. the fact that they belong to the same family or that
they have been observed together. One of the most famous social network of
animals was deeply described in \cite{lusseau-2004-271} and is a
network of social relationships among dolphins: two dolphins are
connected if they have been seen swimming together.

\begin{figure}[!htbp]
  
  \begin{center}
    \includegraphics[scale=0.4]{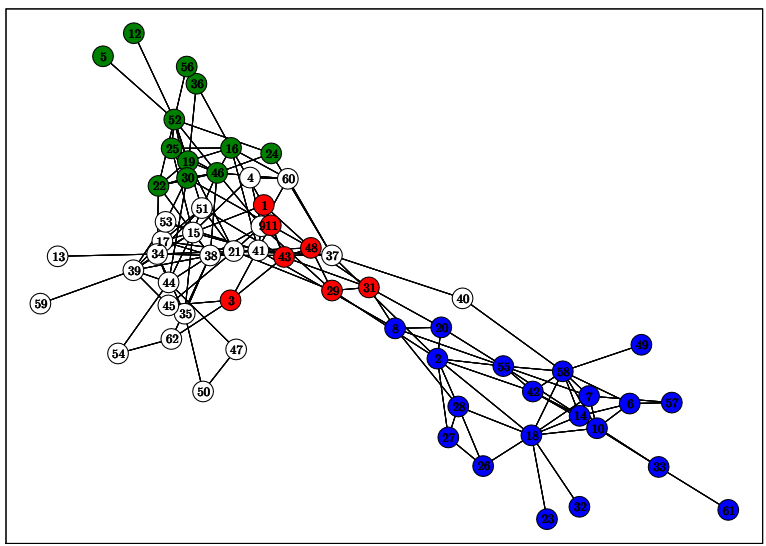}
  \end{center}
  \caption{The four sharp communities found in the dolphins network}
  \label{fig:dolphins4}
\end{figure}

The (perhaps not so) surprising property of the resulting network is that at
least two major communities exist: the first one is formed by females,
while the other contains only males individuals. Nevertheless, it has
been discovered that both those communities are further divided into
sub--communities, and the resulting picture is shown in figure
\ref{fig:dolphins4}.

White and green nodes represent females, while red and blue are for 
males. Note that all females are correctly grouped together as are 
males, and that sub--communities (i.e. green and red nodes) lie
somewhere in the middle between the two main communities (i.e. white
and blue nodes).

The results obtained maximizing the generalized modularity of the
dolphins network are similar to those observed for the Zachary Karate
Club: the four communities are successfully found, and a slight
overlap among sub--communities of each major community is found. For
example, almost all nodes in the small red sub--community of males
have a small overlap with the blue community, and the same happens for
nodes in the green sub--community of females.

\subsection{Political Books}

The problem of the two networks examined so far is that no information
about existing overlaps among communities is available. In the case of
the Zachary network, it is unfeasible to recover such information
since it would be very difficult to find all the people involved in
the experiment and ask them if the overlap found for nodes 3 and 10 is
meaningful, or whether the two sub--communities discovered really were strong
subgroups. In the case of dolphins network, it is simply impossible to
state which overlap makes sense, and which sub--communities have any
real meaning, since this network is based on correlation of
observations made, and not on what animals have ever said about their
real relationships!

On the other hand, information about overlaps in groups and
communities is not given for the majority of networks, and should be
inferred by other characteristics of each network. 

\begin{figure}
  \begin{center}
    \includegraphics[scale=0.3]{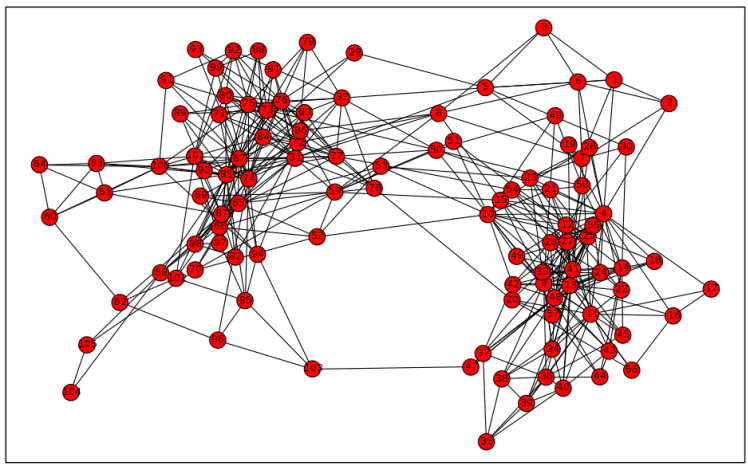}
  \end{center}
  \caption{The polbooks network}
  \label{fig:polbooks_net}
\end{figure}

A complex network which indirectly gives information about overlaps
among communities is the so--called ``PolBooks'' network (see
figure~\ref{fig:polbooks_net}. This network
is an information network where each node represents a political book
sold by Amazon (most of them have been published and edited in the
U.S.) and a link between two nodes exists if a book has been purchased
in combination with another book on the same topic.

Additionally, each node in the network has been marked as ``liberal'',
``conservative'' or ``neutral'', according to the reviews written by
readers. We expect that this network would be naturally divided into
three communities, but the majority of algorithms for community
detection find the best partition with two or four communities, where
nodes labelled as ``neutral'' are often put into the ``conservative''
or ``liberal'' party group. 

\begin{figure}
  \begin{center}
    \includegraphics[scale=0.3]{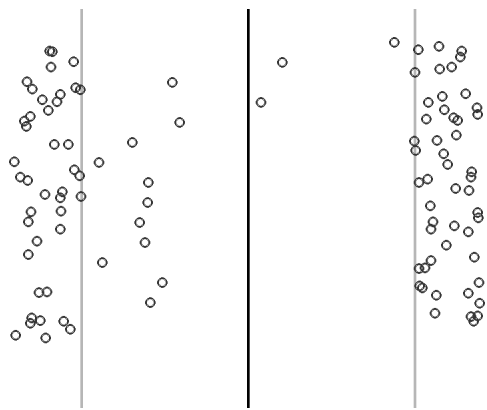}
  \end{center}
  \caption{Optimal partition for the polbooks network}
  \label{fig:polbooks}
\end{figure}

Another image emerges if we maximize the generalized modularity of
the PolBooks network: the best decomposition is that with two
overlapped communities, where conservative and liberal books represent
the larger communities and neutral books are placed as overlapped
among the two, with a small number of books labelled as liberal also
overlapped with the conservative community. Figure \ref{fig:polbooks}
shows this. Note that all the conservative books (nodes on the
right) form a strong community while a modest number of liberal books
show an overlap between the two communities. All books labelled as
neutral are in the middle, and confirm that in such networks the
best decomposition into communities is not sharp and precise, and that
overlaps among communities catch the internal organisation of
the network better.

\subsection{Students in computer engineering}

The last example given in this section is a novel network, built with
the precise objective of testing generalised modularity on graphs
which are intrinsically divided into overlapping communities. 

The network is composed by students of computer engineering course at
the University of Catania: all students attend the third year, and they
are divided into two class, depending on the first letter of their
last name, so that all students whose last name starts with a letter
from A to L is in the first class, while other students are in the
second class. Each class has the same academic curriculum, but
attends lessons with different teachers, and no common courses are
attended by all students together, mainly because of space
limitations. 

Students were asked, through an anonymous test, which one of their
colleagues they knew the best and considered as friends, both in their
own class and in the other one. Each student was assigned a unique ID,
so that other students should use just IDs to answer the question, in
order to preserve the privacy of students. The only personal
information known is the first letter of the name corresponding to
each ID, so that it is possible to label each student as belonging to
the first or to the second class.

It is easy to imagine that the network obtained by connecting nodes
which relate to friends should have communities, since it is much
easier for students to have friends in their own class than in the
other, and the two communities should be almost sharply separated,
since no academic overlap exists among the two groups.

\begin{figure}[!htbp]
  \begin{center}
    \includegraphics[scale=0.3]{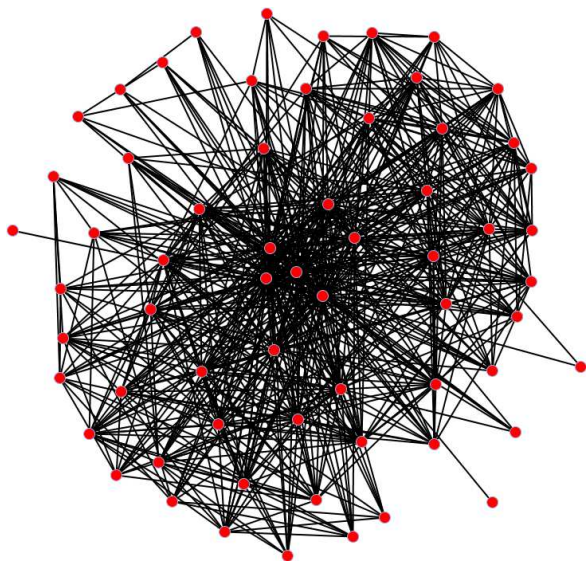}
  \end{center}
  \caption{Engineering students network}
  \label{fig:studenti_net}
\end{figure}

\begin{figure}[!htbp]
  
  \begin{center}
    \includegraphics[scale=0.3]{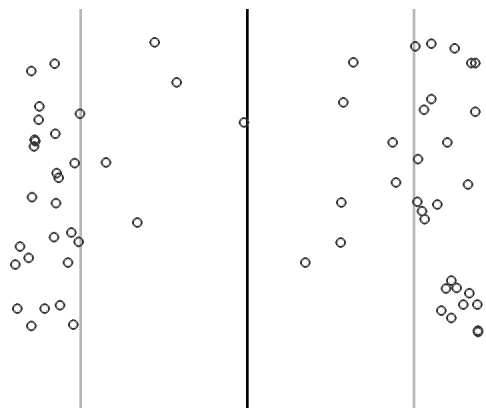}
  \end{center}
  \caption{Overlaps among the two communities in the engineering
    students network}
  \label{fig:studenti}
\end{figure}

Nevertheless, a really surprising result came out of the experiment:
the resulting network (see figure~\ref{fig:studenti_net}), 
which has about sixty nodes and more than one
hundred and fifty links, is quite well connected, and a really low
average path length has been observed. On the other hand, the network
has a clear community structure, but the best decomposition is
obtained when about one--third of the nodes are overlapped among the
two communities. This result is shown in figure \ref{fig:studenti}.

The fact that two groups of students with no common academic
activities can have such a massive overlap sounds strange and somewhat
unusual. We tried to find possible causes of this overlap, but nothing
seemed to can explain it in a crystal clear way. 

Suddenly, we remembered what could seem a secondary detail: all the
students involved in the experiment, who where attending the third
year courses divided into two classes, had been previously divided into
\textit{three} classes during their first year at university, each
class containing students whose name started with a
letter from A to F, from G to P, and from Q to Z, because they
were too many to fit in just two classes! This is the explanation of the
mysterious overlap: the two communities of students were originally
\textit{three} communities and, when the community in the middle was
separated and absorbed by the two final communities, some of the
relationships among students put in two different classes survived, so
the resulting network naturally has communities with massive
overlaps.  Optimization of generalized modularity rediscovered a
forgotten property of the network, finding once again the best
solution. 

The network of students could be used, in the future, as a reference
benchmark for testing capabilities of algorithms for detection of
overlapping communities in complex networks. 

\section{Conclusions} \label{sec:conclusions}

This paper proposed an extension of modularity function for directed
graphs with overlapping communities. This generalization moves from
simple considerations about the meaning and structure of the original
modularity function, using an enriched null--model which takes into
account nodes belonging to more than just one community at the same
time. The function used to estimate the contribute of an edge to the
modularity of a community does not affect the formal procedure used to
derive the generalized modularity, and has been left apart for a future
research. Moreover, a method for overlapping communities discovery based on
the use of a genetic algorithm for the optimization of the extended modularity
function is presented. Finally, we discuss results of the application of our
proposal to several complex networks.

\ack{The authors gratefully acknowledge Santo Fortunato for a careful reading of
the manuscript and for helpful suggestions.  
}

\section*{References}

\bibliographystyle{plain}
\bibliography{finalbib}

\end{document}